\documentstyle[preprint,prb,aps]{revtex}

\tightenlines

\input mathfam

\newmathfam{oca}{msbm10 scaled \magstephalf}{msbm7 scaled
\magstephalf}{msbm5 scaled \magstephalf}{}

\newmathfam{frk}{eufm10 scaled \magstephalf}{eufm7 scaled
\magstephalf}{eufm5 scaled \magstephalf}{}

\def\R{{\oca R}}

\def\l{\left}
\def\r{\right}

\def\lPB{\left\{}
\def\rPB{\right\}}
\def\lpb{\left[\,}
\def\rpb{\,\right]}
\def\com{\,{\mathchar"213B}\,}
\def\xv{{\bf x}}
\def\LieA{{\frk g}}
\def\LieAx{{\frk h}}
\def\LieAxb{{\frk v}}

\def\fd{\delta}
\def\sf{\phi}

\begin{document}

\draft

\title{Invariants and Labels\\ in Lie--Poisson Systems}

\bibliographystyle{pfa}

\author{Jean-Luc Thiffeault%
	\thanks{e-mail: {\tt jeanluc@physics.utexas.edu},\ \
	Tel.: (512)471-6121,\ \ Fax: (512)471-6715}
	and P. J. Morrison%
	\thanks{e-mail: {\tt morrison@hagar.ph.utexas.edu}} }

\address{Institute for Fusion Studies and Department of Physics,\\ University
of Texas at Austin, Austin, Texas, 78712-1060}


\maketitle


\begin{abstract}
Reduction is a process that uses symmetry to lower the order of a Hamiltonian
system.  The new variables in the reduced picture are often not canonical:
there are no clear variables representing positions and momenta, and the
Poisson bracket obtained is not of the canonical type.  Specifically, we give
two examples that give rise to brackets of the noncanonical Lie--Poisson form:
the rigid body and the two-dimensional ideal fluid.  From these simple cases,
we then use the {\it semidirect product} extension of algebras to describe
more complex physical systems.  The Casimir invariants in these systems are
examined, and some are shown to be linked to the recovery of information about
the configuration of the system.  We discuss a case in which the extension is
not a semidirect product, namely compressible reduced MHD, and find for this
case that the Casimir invariants lend partial information about the
configuration of the system.

\end{abstract}

\newpage

This paper explores the Casimir invariants of Lie--Poisson brackets, which
generate the dynamics of some discrete and continuous Hamiltonian systems.
Lie--Poisson brackets are a type of noncanonical Poisson bracket and are
ubiquitous in the reduction of canonical Hamiltonian systems with symmetry.
Casimir invariants are constants of motion for all Hamiltonians; they are
associated with the degeneracy of noncanonical Poisson brackets.
Finite-dimensional examples of systems described by Lie--Poisson brackets
include the heavy top and the moment reduction of the Kida vortex, while
infinite-dimensional examples include the 2--D ideal fluid, reduced
magnetohydrodynamics (MHD), and the 1--D Vlasov equation. (See
Ref.~\onlinecite{Morrison1998} and references therein for a full review\@.)
The Casimir invariants determine the manifold on which the system is
kinematically constrained to evolve.  Understanding the nature of these
constraints is thus of paramount importance.

In Section~\ref{sec:reduction} we examine specific Lie--Poisson brackets,
namely, those that arise from the reduction to Eulerian variables of a
Lagrangian system with relabeling symmetry.  We make use of two prototypical
examples, the rigid body (finite-dimensional) and the 2--D ideal fluid
(infinite-dimensional), and we interpret their Casimir invariants.  This is
done to motivate the introduction of such brackets and to show their physical
relevance.  In Section~\ref{sec:semidirect} we turn to building Lie--Poisson
brackets directly from Lie algebras by the procedure of extension.  We
introduce the semidirect product extension and illustrate it with two physical
examples: the heavy top and low-beta reduced MHD.  Finally in
Section~\ref{sec:CRMHD} we look at a nonsemidirect example, compressible
reduced MHD, and discuss work in progress.

\section{Lie--Poisson Brackets and Reduction}
\label{sec:reduction}

For our purposes, a reduction is a mapping of the dynamical variables of a
system to a smaller set of variables,  such that the transformed
Hamiltonian and bracket depend only on the  smaller set of variables.  (See for
example Ref.~\onlinecite{MarsdenRatiu} for a detailed treatment\@.)  The
simplest example of a reduction is the case in which a cyclic variable is
eliminated, but more generally  a reduction exists as a consequence  of an
underlying symmetry of the system.  We present two examples of reduction.

\subsection{Reduction of the Free Rigid Body}
\label{sec:reductionrigid}

The Hamiltonian for the free rigid body is an unwieldy function of three Euler
angles $\phi,\psi,\theta$ and their conjugate momenta
$p_\phi,p_\psi,p_\theta$.
The motion is described by Hamilton's equations using the canonical bracket
\[
\l\{f\com g\r\}_{\rm can} =
\frac{\partial f}{\partial\phi}\,\frac{\partial g}{\partial p_\phi}
-\frac{\partial g}{\partial\phi}\,\frac{\partial f}{\partial p_\phi}
+\frac{\partial f}{\partial\psi}\,\frac{\partial g}{\partial p_\psi}
- \frac{\partial g}{\partial\psi}\,\frac{\partial f}{\partial p_\psi}
+\frac{\partial f}{\partial\theta}\,\frac{\partial g}{\partial p_\theta}
- \frac{\partial g}{\partial\theta}\,\frac{\partial f}{\partial p_\theta}.
\]
Here we have 3 degrees of freedom (6 coordinates), the configuration space is
the rotation group $SO(3)$, and the phase space is its contangent
bundle~$T^*SO(3)$.

A reduction is possible for this system.  In terms of angular
momenta $\ell_i$ about the principal axes, we have
\[
	H(\phi,\psi,\theta,p_\phi,p_\psi,p_\theta)\ \longrightarrow\
	H(\ell_1,\ell_2,\ell_3) = \sum_{i=1}^3 \frac{\ell_i^2}{2I_i}\,.
\]
Under this (noncanonical) mapping, the bracket obtains the Lie--Poisson
form
\[
\l\{f\com g\r\} = -\ell\cdot\,\frac{\partial f}{\partial \ell}
	\times\frac{\partial g}{\partial \ell}\,.
\]
The equations of motion generated by this bracket and
$H(\ell_1,\ell_2,\ell_3)$ are Euler's equations  for the rigid body,
\[
	\dot\ell_i = \frac{I_j-I_k}{I_j\,I_k}\,\ell_j\,\ell_k\,,
\]
where $i,j,k$ are cyclic permutations of $1,2,3$.   The energy is
conserved, and so is the quantity
\[
	C(\ell) = \sum_{i=1}^3 \ell_i^2\,,
\]
which commutes with any function of $\ell$.  Such a functions are called
Casimir invariants (or Casimirs for short).  Casimirs are conserved quantities
for any Hamiltonian, so they tell us about the topology of the manifold on
which the motion takes place.  For the simple case of the rigid body, the
motion takes place on the two-sphere, $S^2$ (not in physical space, but in
angular momentum space).  The symmetry that permits the reduction is the
invariance of the equations of motion for
$(\phi,\psi,\theta,p_\phi,p_\psi,p_\theta)$ under rotations (elements
of~$SO(3)$).  This symmetry amounts to the freedom of choosing axes from which
the Euler angles are measured.  In that sense it is a relabeling symmetry,
since the choice of axes amounts to making ``marks,'' or labels, on the rigid
body.

We shall say that the original system is a Lagrangian description (by analogy
with the fluid case below) because at any time the exact configuration of the
system (including orientation) is known, whereas the reduced system is
Eulerian because only the angular momentum of the body is known.

\subsection{Reduction of the 2--D Ideal Fluid}
\label{sec:red2dfl}

As our prototype for reduction in an infinite dimensional system we take an
ideal 2--D fluid confined to some domain, $D$.  The Lagrangian description
involves fluid elements labeled by some coordinate $a$, which is usually taken
to be the initial position of the fluid elements.  These labels are analogous
to the choice of axes in the rigid body example above (the ``marks'' on the
rigid body).  The Hamiltonian functional is
\[
	H[q;\pi] = \int_D\,\l\lgroup\frac{\pi^2}{2\rho_0}
		- p(a,t)
		\l(\,\l|\frac{\partial q}{\partial a}\r|-1\r)\r\rgroup\,d^2a,
\]
where $q(a,t)$ is the position of the fluid element labeled by $a$ and
$\pi(a,t)$ is its momentum.  The Jacobian of the transformation from the
labels~$a$ to the position of the fluid elements at a later time
is~$|{\partial q}/{\partial a}|$.  The density $\rho_0$ is taken to be
constant, and the pressure~$p(a,t)$ appears here as a Lagrange multiplier
that enforces the incompressibility condition,~$|{\partial q}/{\partial
a}|=1$ (see Ref.~\onlinecite{Newcomb1962}).  This Hamiltonian together with
the canonical bracket
\[
	{\{F\com G\}}_{\rm can} = \int_D \l\lgroup\frac{\delta F}{\delta q}\,
		\frac{\delta G}{\delta \pi} - \frac{\delta G}{\delta q}\,
		\frac{\delta F}{\delta \pi}\r\rgroup\,d^2a
\]
generates the equations of motion for a fluid in Lagrangian variables.  The
information about the position of every fluid element at any time is contained
in the model. The dynamical evolution of the system is independent of the
particular choice of labels for the fluid elements.  This {\it relabeling}
symmetry of the initial condition labels, $a$, suggests a reduction.

We introduce the streamfunction $\sf$ defined by $v(\xv,t) =
(-\partial_y\,\sf,\partial_x\,\sf)$, so that $\nabla\cdot v=0$ is
automatically satisfied, and the vorticity $\omega(\xv,t) = \nabla^2\sf$.  The
mapping from the Lagrangian momentum to the Eulerian velocity field is
\[
	{\rho_0}v(\xv,t) = \int_D\,{\pi(a,t)}\,
		\delta(\xv-q(a,t))\,d^2a\,.
\]
We take~$\rho_0=1$ for simplicity.  Taking the curl of~$v$ and dotting
with~$\hat z$ gives
\[
	\omega(\xv,t)=\hat z \cdot \nabla\times v(\xv,t) =
		\int_D\,\epsilon_{ij}\,\pi_i(a,t)\,
		\partial_j\delta(\xv-q(a,t))\,d^2a\,,
\]
where repeated indices are summed,~$\partial_i := \partial/\partial x_i$, and
the antisymmetric symbol~$\epsilon_{ij}$ is defined by~$\epsilon_{12}=1$.  The
variation of~$\omega$ is
\[
	\delta \omega(\xv,t)=\int_D\,\l(\epsilon_{ij}\,\delta\pi_i(a,t)\,
		\partial_j\delta(\xv-q(a,t)) -
		\epsilon_{ij}\,\pi_i(a,t)\,
		\partial_j\partial_k\delta(\xv-q(a,t))
		\,\delta q_k(a,t)\r)\,d^2a\,,
\]
which we can insert into
\[
	\delta F = \int_D\,\frac{\delta F}{\delta\omega}
		\,\delta\omega\,d^2x
	= \int_D\,\l(\frac{\delta F}{\delta \pi}
		\,\delta \pi+\frac{\delta F}{\delta q}
		\,\delta q\r)d^2a
\]
to find
\begin{eqnarray*}
	\frac{\delta F}{\delta \pi} &=& {\l.\hat z \times
		\nabla\l(\frac{\delta F}{\delta \omega}\r)\r|
		}_{\xv=q(a,t)},\\
	\frac{\delta F}{\delta q} &=& -\int_D\,\epsilon_{ij}\,
		\pi_i(a,t)\,\partial_j\nabla\delta(\xv-q(a,t))
		\,\frac{\delta F}{\delta \omega}\,d^2x.
\end{eqnarray*}
We then insert these two expressions into the canonical bracket.  After some
manipulation involving integration by parts (we assume boundary terms vanish)
we obtain the Lie--Poisson bracket
\[
	\{F\com G\} = \int_D \omega\,\l[\frac{\delta F}{\delta \omega}\com
		\frac{\delta G}{\delta \omega}\r]\,d^2x\,,
\]
where
\[
	[f\com g] :=
	\frac{\partial f}{\partial x}\,\frac{\partial g}{\partial y}
	- \frac{\partial f}{\partial y}\,\frac{\partial g}{\partial x}\,.
\]
Note that the incompressibility of the fluid (the fact that the
Jacobian~$|\partial q/\partial a|$ is unity) was not used in the derivation of
this bracket.  However, in order to write the Hamiltonian in terms of~$\omega$
one must introduce the streamfunction~$\sf$, which is possible only
if~$\nabla\cdot v=0$.  The equation of motion generated by the bracket and the
transformed Hamiltonian
\[
	H[\omega] = -\frac{1}{2}\int_D\,\sf\,\omega\,\, d^2x
		= \frac{1}{2}\int_D\,|\nabla\sf|^2\,\, d^2x
\]
is just Euler's equation for an the ideal fluid
\[
	\dot\omega(\xv) = -[\sf\com\omega]\,.
\]
This has a Casimir given by
\[
	C[\omega] = \int_D\,f\l(\omega(\xv)\r)\,d^2x,
\]
where~$f$ is an arbitrary function.  The interpretation of this invariant is
given in detail in Ref.~\onlinecite{Morrison1987}.  It implies the
preservation of contours of~$\omega$, so that the value~$\omega_0$ on a
contour labels that contour for all times.  This is a consequence of the
dissipationless and divergence-free nature of the system.  Substituting
$f(\omega) = \omega^n$ we also see that all the moments of vorticity are
conserved.  By choosing~$f(\omega) = \theta(\omega(\xv) - \omega_0)$, a
heavyside function, it follows that the area inside of any
$\omega$-contour is conserved.

\section{Extensions and the Semidirect Product}
\label{sec:semidirect}

We now investigate systems involving Lie algebras by {\it extension}, a
procedure for combining two or more Lie algebras to make a new Lie algebra.
There are a myriad of ways to extend algebras, and we will only touch on a few
here.  All the extensions discussed here have their equivalent for Lie groups,
but we choose the algebra approach here because it leads more directly to a
Lie--Poisson bracket.  In this section we let~$\LieAx$ be an extension of
the Lie algebra~$\LieA$ by the algebra~$\LieAxb$. The elements of~$\LieAx$ are
written as 2-tuples,~$(\xi,\eta)$, where~$\xi\in \LieA$ and~$\eta\in \LieAxb$.

The simplest extension is the {\it direct product} of Lie algebras.  Let $\xi$
and $\xi'$ be elements of a Lie algebra $\LieA$ and $\eta$ and $\eta'$ be
elements of a vector space $\LieAxb$ (which is an Abelian Lie algebra under
addition).  The direct product of these two algebras is an algebra $\LieAx$
with bracket
\[
\lpb(\xi\com\eta)\com(\xi'\com\eta')\rpb :=
	\l(\lpb\xi\com\xi'\rpb\com\lpb\eta\com\eta'\rpb\r).
\]

Given the same $\LieA$ and $\LieAxb$ as above there is a less trivial way to
make a new Lie algebra called the {\it semidirect product} with an operation
defined by
\[
\lpb(\xi\com\eta)\com(\xi'\com\eta')\rpb :=
	\l(\lpb\xi\com\xi'\rpb\com
	\lpb\xi\com\eta'\rpb + \lpb\eta\com\xi'\rpb\r).
\]
A simple example of a semidirect product structure is when $\LieA$ is the Lie
algebra~$so(3)$ associated with the rotation group $SO(3)$ and $\LieAxb$ is
${\oca R}^3$.  Their semidirect product is the algebra of the 6-parameter
Galilean group of rotations and translations.  Both the elements of~$\LieA$
and~$\LieAxb$ can then be represented by vectors in~$\R^3$, with
bracket~$\lpb\xi\com\eta\rpb=\xi\times\eta$, the cross product of vectors.
Since~$\LieAx$ is itself a Lie algebra, it can be extended again as needed to
make an $n$-fold extension (an algebra of $n$-tuples).

We can build Lie--Poisson brackets from these algebras by
extension.\cite{Marsden1984}  For an $n$-fold extension~$\LieAx$ of the Lie
algebra~$\LieA$, we define
\[
\lPB F\com G\rPB := \pm\l\langle\mu\com\lpb
	\frac{\delta F}{\delta \mu}\com\frac{\delta G}{\delta \mu}
	\rpb\r\rangle,
\]
where~$\mu\in \LieAx^*$, the dual of~$\LieAx$ under the
pairing~$\langle\com\rangle:\LieAx^*\times\LieAx\rightarrow\R$.  The dynamical
variables of the system are the elements of the~$n$-tuple~$\mu=\mu(t)$.  These
elements may be fields or variables, so the Lie--Poisson bracket derived from
an algebra by extension generates the dynamics for a system involving several
dynamical quantities.  The functions (or functionals) $F$ and $G$ are maps
from~$\LieAx^*$ to~$\R$.  Here~$\delta/\delta\mu$ is a derivative or a
functional derivative, depending on the dimensionality of the algebra (finite
or infinite).  For~$n=1$,~$\LieAx=so(3)$, we have~$\mu:=\ell$ and we recover
the bracket for the free rigid body (Section~\ref{sec:reductionrigid}).  The
overall sign of the Lie--Poisson bracket has to do with left- or
right-invariance of vector fields and will not be discussed here (See
Ref.~\onlinecite{MarsdenRatiu}).

Using this procedure to make a Lie--Poisson bracket from a direct product of
algebras leads to a sum of~$n$ independent brackets.  This will not interest
us further, since the coupling between dynamical variables can only come from
the Hamiltonian.  However there are interesting physical examples of a direct
product structure.  (This is the case for the model in
Ref.~\onlinecite{Hazeltine1985}, although a coordinate transformation is
needed to exhibit the structure.)

We illustrate the process of building a Lie--Poisson bracket from a semidirect
product of algebras by two examples, which are extensions of the rigid body
and ideal fluid examples of Section~\ref{sec:reduction}.

\subsection{The Heavy Top}

The Lie--Poisson bracket for the semidirect product of the rotation group
$SO(3)$ and the vector space ${\oca R}^3$ is
\[
\l\{f\com g\r\} = -\ell\cdot\l(\frac{\partial f}{\partial \ell}
	\times\frac{\partial g}{\partial \ell}\r)
	- \alpha\cdot\l(\frac{\partial f}{\partial \ell}
		\times\frac{\partial g}{\partial \alpha}
		+ \frac{\partial f}{\partial \alpha}
		\times\frac{\partial g}{\partial \ell}\r)
\]
where $\alpha$ denotes a 3-vector.  By using
\[
	H(\ell,\alpha) = \sum_{i=1}^3\frac{\ell_i^2}{2I_i}+\alpha\cdot{\bf c}
\]
where~${\bf c}$ is a vector representing the position of the center-of-mass,
we get the prototypical example of a semidirect product system, the heavy
rigid body (in the body frame):
\begin{eqnarray*}
	\dot\ell_i = \frac{I_j-I_k}{I_j\,I_k}\,\ell_j\,\ell_k
		+ \,\alpha_j\,c_k - \,\alpha_k\,c_j\,,\ \ \ \
	\dot\alpha_i = \frac{\ell_k\,\alpha_j}{I_k}
		- \frac{\ell_j\,\alpha_k}{I_j}\,,
\end{eqnarray*}
where $i,j,k$ are cyclic permutations of $1,2,3$. The vector $\alpha$ rotates
rigidly with the body, which is always true for a Hamiltonian quadratic in
$\ell$.  The Casimirs for this bracket are
\[
	C_1 = \alpha^2\,,
	\ \ \ C_2 = \ell\cdot\alpha\ .
\]
Looking at the bracket as derived by reduction of the heavy top in Euler
angles (as we did in Section~\ref{sec:reductionrigid}, but here with gravity),
the Casimir $C_2$ expresses conservation of $p_\phi$, since $\phi$ is cyclic.
Knowing $\alpha$ does not lead to a determination of the orientation of the
rigid body: there is still a symmetry of rotation about $\alpha$.  Taking the
semidirect product has led to the recovery of some of the Lagrangian
(configuration) information.

\subsection{Low-beta Reduced MHD}
\label{sec:lbrmhd}

The semidirect product bracket for two fields $\omega$ and $\psi$ is
\begin{eqnarray*}
	\{F\com G\} &=& \int_D \l\lgroup\omega\l[\frac{\delta F}{\delta \omega}
			\com\frac{\delta G}{\delta \omega}\r]
	+ \psi\l(\l[\frac{\delta F}{\delta \omega}
			\com\frac{\delta G}{\delta \psi}\r]
			+ \l[\frac{\delta F}{\delta \psi}
			\com\frac{\delta G}{\delta \omega}\r]\r)\r\rgroup
	d^2x\,.
\end{eqnarray*}
If $\omega=\nabla^2\sf$, where $\sf$ is the electric potential, $\psi$ is
the magnetic flux, and $J=\nabla^2\psi$ is the current, then the Hamiltonian
\[
	H[\omega;\psi] = \frac{1}{2}\int_D\,\l(
		|\nabla\sf|^2+|\nabla\psi|^2\r)\,d^2x
\]
with the above bracket gives us
\begin{eqnarray*}
	\dot\omega = \l[\omega,\sf\r] + \l[\psi,J\r] \ ,\ \ \ \
	\dot\psi = \l[\psi,\sf\r]\ ,
\end{eqnarray*}
a model for low-beta reduced
MHD.\cite{Morrison1984} (Ref. \onlinecite{Benjamin1984} contains a system
with a similar structure, but for waves in a density-stratified fluid.)

The bracket has two Casimir invariants,
\[
	C_1[\psi] = \int_D\,f(\psi)\,d^2x,\ \ \ \
	C_2[\omega;\psi] = \int_D\,\omega\,g(\psi)\,d^2x.
\]
The first has the form of the Casimir for 2--D Euler of
Section~\ref{sec:red2dfl} and has the same interpretation.
To understand the second one we let $g(\psi)=\theta(\psi-\psi_0)$, a
heavyside function.  In this case we have
\[
	C_2[\omega;\psi] = \int_{\Psi_0}\,\omega\,d^2x,
\]
where~$\Psi_0$ represents the (not necessarily connected) region of $D$
enclosed
by the
contour $\psi=\psi_0$, and~$\partial\Psi_0$ is its boundary.  The
contour $\partial\Psi_0$ moves with the fluid, so this just expresses Kelvin's
circulation theorem: the circulation around a closed material loop is
conserved.

This theorem is true for any $\psi$-contour, therefore it holds in the region
between two contours $\psi=\psi_0$ and $\psi=\psi_0+\delta$.  Letting
$\delta\rightarrow 0$ we see that the two contours delineate a ``line'' of
fluid elements with value~$\psi_0$ of the magnetic flux.  Knowledge of the
value of~$\psi$ on a fluid element thus only determines which contour it is
on, but not its location on the contour.  Therefore, there is still a
relabeling symmetry: the fluid elements can be shifted around the contour
without changing the Casimirs~$C_1$ and~$C_2$.  As with the heavy top, the
semidirect product has led to the recovery of some, but not all, of the
Lagrangian information.


\subsection{Putting Labels on a Rigid-body}

Remember that taking a semidirect product restricted the symmetry group of the
body to rotations about $\alpha$.  If we take another semidirect product to
get
\begin{eqnarray*}
\l\{f\com g\r\} = -\ell\cdot\l(\frac{\partial f}{\partial \ell}
	\times\frac{\partial g}{\partial \ell}\r)
	- \alpha\cdot\l(\frac{\partial f}{\partial \ell}
		\times\frac{\partial g}{\partial \alpha}
		+ \frac{\partial f}{\partial \alpha}
		\times\frac{\partial g}{\partial \ell}\r)
	- \beta\cdot\l(\frac{\partial f}{\partial \ell}
		\times\frac{\partial g}{\partial \beta}
		+ \frac{\partial f}{\partial \beta}
		\times\frac{\partial g}{\partial \ell}\r)
\end{eqnarray*}
where $\beta$ is a 3-vector, we have a bracket that can model a rigid body with
two forces acting on it, for example a charged, rigid insulator in an electric
field.  The new bracket has Casimirs
\[
	C_1 = \alpha^2\,,
	\ \ \ C_2 = \beta^2,
	\ \ \ C_3 = \alpha\cdot\beta\,.
\]
The angular momentum $\ell$ has disappeared from the Casimirs.  This is
because knowing $\alpha$ and $\beta$ completely specifies the orientation of
the rigid body (unless the two are colinear).  In other words, by taking
semidirect products we have reintroduced the Lagrangian information into the
bracket.  Note that taking more than two semidirect products is redundant as
far as the Lagrangian information is concerned: knowing the orientation of
more than two vectors does not add new information.  This is reflected by the
fact that the number of variables minus the number of Casimirs is six for
two or more ``advected'' quantities.  This is the dimension of~$T^*SO(3)$, the
original phase space (before reduction).

\subsection{Advection in an Ideal Fluid}

We now take a second semidirect product for the ideal fluid, say low-beta
MHD with a second advected quantity, the pressure~$p$.  In that case we get a
model for high-beta reduced MHD.\cite{Strauss1977}  The Casimir is
\[
	C[\psi;p] = \int_D\,f(\psi,p)\,d^2x,
	\ \ \ f\ {\rm arbitrary}.
\]
This Casimir amounts to being able to label two contours.  Locally this
permits a unique labeling of the fluid elements as long as $p$ and $\psi$ are
not constant in some region.  However, globally there is some ambiguity,
because contours can cross in several places.  Thus, in the
infinite-dimensional case the semidirect product is not equivalent to
recovering the full Lagrangian information, unless the contours do not close,
are monotonic, and nonparallel (\hbox{$\nabla\psi \times \nabla p$} does not
vanish).  A third advected quantity will in general break this degeneracy.
Note that if the advected quantities label the fluid elements unambiguously at
$t=0$ then they will do so for all times.

\section{Beyond the Semidirect Product: Cocycles}
\label{sec:CRMHD}

In general there are other ways to extend Lie algebras besides the semidirect
product.
One example is the model derived in
Refs.~\onlinecite{Hazeltine1985b} and \onlinecite{Hazeltine1987} for 2--D
compressible reduced MHD.  The model has four fields, and is obtained from an
expansion in the inverse aspect ratio of the tokamak.  The Hamiltonian is
\[
	H[\omega,v,p,\sf] = \frac{1}{2}\int_D\l( |\nabla\sf|^2 + v^2
		+ \frac{(p-2\beta\,x)^2}{\beta} + |\nabla\psi|^2\r) \,d^2x,
\]
where $v$ is the parallel velocity, $p$ is the pressure, and $\beta$ is a
parameter that measures compressibility.  The bracket is
\begin{eqnarray*}
	\lPB A\com B\rPB &=&
		\int_D\, d^2x\l\lgroup\omega\lpb\frac{\fd A}{\fd \omega}
			\com\frac{\fd B}{\fd \omega}\rpb\r.\\
	&&\ \ \ \ \ \mbox{}+ v\l(\lpb\frac{\fd A}{\fd \omega}
			\com\frac{\fd B}{\fd v}\rpb
		+ \lpb\frac{\fd A}{\fd v}
			\com\frac{\fd B}{\fd \omega}\rpb\r)
	+ p\l(\lpb\frac{\fd A}{\fd \omega}
			\com\frac{\fd B}{\fd p}\rpb
		+ \lpb\frac{\fd A}{\fd p}
			\com\frac{\fd B}{\fd \omega}\rpb\r)\\
	&&\ \ \ \ \ \mbox{} + \l.\psi\l(\lpb\frac{\fd A}{\fd \omega}
			\com\frac{\fd B}{\fd \psi}\rpb
		+ \lpb\frac{\fd A}{\fd \psi}
			\com\frac{\fd B}{\fd \omega}\rpb
	- \beta \lpb\frac{\fd A}{\fd p}
			\com\frac{\fd B}{\fd v}\rpb
		-\beta \lpb\frac{\fd A}{\fd v}
			\com\frac{\fd B}{\fd p}\rpb\r)\r\rgroup.
\end{eqnarray*}
The term proportional to $\beta$ is an obstruction to the semidirect product
structure, and it cannot be removed by a coordinate transformation.

The theory that deals with the classification of extensions is Lie algebra
cohomology.  In general the way to extend a bracket is by adding a nontrivial
{\it cocycle}.  Though a priori there are an infinite number of ways to make
an extension, for low dimensions, after allowing for coordinate
transformations, very few possibilities remain;  we have classified these in
Ref.~\onlinecite{JLTinprep}.  We have also found all the Casimir invariants
for the low-dimensional brackets (five fields or less).

The Casimirs of the above bracket are
\begin{eqnarray*}
	&&C_1[\psi] = \int_D\,f(\psi)\,d^2x,\ \ \ \ \ \ \ \ \
	C_2[p;\psi] = \int_D\,p\, g(\psi)\,d^2x,\\
	&&C_3[v;\psi] = \int_D\,v\, h(\psi)\,d^2x,\ \ \ \
	C_4[\omega,v,p,\psi] = \int_D\,\l(\omega\, k(\psi)
		+ \frac{v\,p}{\beta}\,k'(\psi)\r)\,d^2x.
\end{eqnarray*}
Finding the invariant $C_4$ directly from the equations of motion would be
tedious, but is  straightforward from the bracket.  These Casimirs
do not allow a labeling of the fluid elements.  The meaning of invariants of
the form of $C_1$, $C_2$, and $C_3$ was discussed in
Sections~\ref{sec:red2dfl} and~\ref{sec:lbrmhd}:  the total magnetic flux,
pressure,  and parallel velocity inside of any $\psi$-contour are preserved.
To understand $C_4$ we use the fact that~$\omega=\nabla^2\sf$ and then
integrate
by parts to obtain
\[
	C_4[\omega,v,p,\psi] = \int_D\,\l(-\nabla\sf\cdot\nabla\psi
		+ \frac{v\,p}{\beta}\r)\,k'(\psi)\,d^2x.
\]
The quantity in parentheses is thus invariant inside of any $\psi$-contour.  It
can be shown that this is a remnant of the conservation in the full MHD model
of the cross helicity,
\[
	V = \int_D {\bf v}\cdot{\bf B}\,d^2x\,,
\]
at second order in the inverse aspect ratio, while  $C_3$ is a consequence of
preservation of this quantity at first order.  Here ${\bf B}$ is the magnetic
field.  As for $C_1$ and $C_2$ they are, respectively, the first and second
order remnants of the preservation of helicity,
\[
	W = \int_D {\bf A}\cdot{\bf B}\,d^2x,
\]
where ${\bf A}$ is the magnetic vector potential.


\section{Conclusions}

We gave an introduction to the reduction of physical systems based on their
symmetries.  The prototypical examples were shown, the rigid body and the 2--D
ideal fluid.  For these two cases some information about the configuration of
the system was lost after reduction, correponding to the symmetry used to
reduce the system.

The semidirect product allowed us to build larger brackets from a ``base''
algebra in a systematic manner.  We were thus able to describe the heavy top
and low-beta reduced MHD.  Examining the invariants, we concluded that the
semidirect product had recovered some or all of the Lagrangian information.

For general extensions (not necessarily semidirect) things are different: the
Lagrangian information is not necessarily a consequence of the Casimirs.
However, for compressible reduced MHD the Casimirs represent constraints that
are remnants of invariants of the full MHD equations from which the model is
derived asymptotically.

As mentioned in Section~\ref{sec:CRMHD}, when considering a general extension,
all brackets can be reduced to a small number of normal forms, at least for
low-dimensional extensions.  It will be interesting to see if physical systems
can be found that are realized by these brackets, both in the finite and
infinite degree-of-freedom cases.  We are currently investigating a toy model
that we call the Leibniz top, which is one of the simplest non-semidirect
system one can build.  It is a straightforward generalization of the Lagrange
top (a heavy top with~$I_1=I_2$).  The Lagrange top is integrable, and we have
found that so is the Leibniz top.

\acknowledgements

This work was supported by the U.S. Dept.~of Energy Contract
No.~DE-FG03-96ER-54346.  J.-L.T. also acknowledges support from the Fonds pour
la Formation de Chercheurs et l'Aide \`a la Recherche du Canada.


\end{document}